\documentclass[%
onecolumn,
showkeys,
showpacs,
nofootinbib,
amsmath,amssymb,
10pt,
longbibliography]{revtex4}

\usepackage[colorlinks,citecolor=blue,urlcolor=blue,linkcolor=blue]{hyperref}
\usepackage{array,multirow,graphicx}
\usepackage[
scale=0.7, marginratio={1:1, 2:3}, ignoreall,
text={7in,12in},centering,
margin=1.5in,
total={6.5in,8.75in}, top=1.2in, left=0.9in, includefoot,
height=10in,a4paper,hmargin={3cm,0.8in},
]{geometry}

\newcommand{\e}[2]{{\mathop{#1}\limits_{#2}}}
\newcommand{\h}[1]{{\mathop{\lambda}\limits_{#1}}}
\newcommand{\s}[1]{{\mathop{#1}\limits^{*}}}
\newcommand{\n}[1]{{\mathop{#1}\limits^{(0)}}}
\newcommand{\nn}[1]{{\mathop{#1}\limits^{(1)}}}
\newcommand{\nnn}[1]{{\mathop{#1}\limits^{(2)}}}

\newcommand\undersym[2]{\raisebox{-6pt}{\tiny$#2$}{\kern-5pt}\mbox{$#1$}}

\begin{document}
\title{A pure geometric theory for gravity and material distribution}
\author{M. I. Wanas$^{1,4}$}%
\email{miwanas@sci.cu.edu.eg}
\author{Nabil L. Youssef$^{2,4}$}
\email{nlyoussef@sci.cu.edu.eg}
\author{W. El Hanafy$^{3,4}$}%
\email{waleed.elhanafy@bue.edu.eg}
\affiliation{$^{1}$Department of Astronomy, Faculty of Science, Cairo University, Egypt}
\affiliation{$^{2}$Department of Mathematics, Faculty of Science, Cairo University, Egypt}
\affiliation{$^{3}$Centre for Theoretical Physics, The British University in Egypt, P.O. Box 43, El Sherouk City, Cairo 11837, Egypt}
\affiliation{$^{4}$Egyptian Relativity Group (ERG), Cairo University, Giza 12613, Egypt}
\begin{abstract}
A field theory is constructed in the context of parameterized absolute parallelism geometry. The theory is shown to be a pure gravity one. It is capable of describing the gravitational field and a material distribution in terms of the geometric structure of the geometry used (the parallelization vector fields). Three tools are used to attribute physical properties to the geometric objects admitted by the theory. Poisson and Laplace equations are obtained in the linearized version of the theory. The spherically symmetric solution of the theory, in free space, is found to coincide with the Schwarzschild exterior solution of the general theory of relativity. The theory respects the weak equivalence principle in free space only. Gravity and material distribution are not minimally coupled.
\end{abstract}

\maketitle
\section{Introduction}\label{S1}
Gravity is one of the known four interactions usually used to interpret different phenomena in our Universe. It is widely accepted that the theory of General Relativity (GR) is the best theory for gravity, so far. The theory is successful in the solar and stellar systems while it has some problems with observations in larger systems, e.g., the rotation velocity of stars in spiral galaxies \cite{BBS91}, observations of SN type Ia \cite{SN98}, ..., etc. In the past three decades, many authors have attempted to modify GR or write new theories for gravity to get rid of such problems. Some authors suggested a type of modification of GR by using an action made of a function of the scalar curvature $R$. This type of modification is known in the literature as $f(R)$-theories (cf. \cite{Nojiri:2006ri, F2010, SF2010, Nojiri:2011pr, Nojiri:2013zza}) and is carried out in the context of Riemannian geometry. Another type of modification of GR is known in the literature as $f(T)$-theories (cf. \cite{BF09, L10, Bamba:2011JCAP, Bamba:2011pz, LSB11, Z2011, G1, Cai:2011tc, Capozziello:2011hj, G2, G3, Nashed:2014lva, CQSW14, Hanafy:2014ica, Hanafy:2014bsa, Saridakis2, Odintsov:2015AnPhy, Hanafy2016, Bamba:2016gbu,Nashed2017}), $T$ being the torsion scalar giving the teleparallel equivalent of general relativity, TEGR, (cf. \cite{A2004, G4}). This class of modification is constructed in the context of absolute parallelism (AP-) geometry.

The geometry used by the first group, Riemannian geometry, is characterized by a symmetric metric linear connection, the Levi-Civita connection. The second group uses a geometry with a non-symmetric linear connection, Absolute Parallelism (AP-) geometry. The connection is metric and has a non-vanishing torsion but vanishing curvature. Both groups are working in the context of the geometerization philosophy. This philosophy has proved to be very successful in describing gravity.

In the present work, we use a geometry which has the advantages of both Riemannian and AP-geometries, the Parametrized Absolute Parallelism (PAP-) geometry. Its linear connection is metric and has simultaneously non-vanishing torsion and non-vanishing curvature. The aim of the present work is to construct a field theory, in the context of PAP-geometry, and to explore its physical consequences.

The article is arranged as follows. In Section \ref{S2}, we review briefly the main features of the PAP-geometry. In Section \ref{S3}, we discuss the choice of a particular Lagrangian and the choice of the field equations as well. In Section \ref{S4}, the field equations have been deduced in terms of the PAP-tensors. The analysis of the field theory has been obtained in Section \ref{S5}, this has been done not only by covering the most successful nonlinear theories but also by comparing the linearized field equations with the successful linear theories. The linearization of the symmetric part of the field equations leads to Poisson and Laplace equations. Moreover, we give a physical classification of the field equations using a scheme known in the literature as the type analysis scheme. In Sections \ref{S6}, we apply the field equations to a spherically symmetric spacetime, the solution perfectly produces the Schwarzschild solution in free space. The work has been concluded by some comments and remarks, Section \ref{S7}.

\section{A Brief Review of PAP-Geometry}\label{S2}

In this section we give first a brief account of the AP-geometry. For more details, we refer for example to \cite{M62, W2001, NW13, NA07, NA08}. Then, we briefly review the PAP-geometry. For more details, we refer to \cite{W98, W2000,Wanas:2016boh}.

An AP-space is a pair $(M,\,\undersym{\lambda}{i})$, where $M$ is an $n$-dimensional smooth manifold and $\undersym{\lambda}{i}$ ($i=1,\cdots, n$) are $n$ independent vector fields defined globally on $M$. The vector fields $\undersym{\lambda}{i}$ are called the parallelization vector fields. An AP-space is also known in the literature as a parallelizable or a teleparallel space.

Let\, $\undersym{\lambda}{i}^{\mu}$ $(\mu = 1, ..., n)$ be the coordinate components of the $i$-th vector field
$\undersym{\lambda}{i}$. The Einstein summation convention is applied on both Latin (mesh) and Greek (world) indices, where
Latin indices are written beneath the symbol $\lambda$. The covariant components \ $\undersym{\lambda}{i}_{\mu}$ of\, $\undersym{\lambda}{i}$ are given via the relations
\begin{equation}\label{orthonormality}
\undersym{\lambda}{i}^{\mu}\,\undersym{\lambda}{i}_\nu = \delta^{\mu}_{\nu}, \ \ \
\undersym{\lambda}{i}^{\mu}\,\undersym{\lambda}{j}_{\mu} =
\delta_{ij}.
\end{equation}
Because of the independence of \,$\undersym{\lambda}{i}$, the determinant $\lambda:=\det (\,\undersym{\lambda}{i}^{\mu})$ is nonzero.

On an AP-space $(M,\,\undersym{\lambda}{i})$, there exists a unique linear connection with respect to which the parallelization vector fields are parallel. This connection is given by
\begin{equation}\label{canonical}
\Gamma^{\alpha}_{~\mu\nu}: =
\undersym{\lambda}{i}^{\alpha}\,\undersym{\lambda}{i}_{\mu,\nu}
\end{equation}
and is characterized by the property that
\begin{equation}\label{AP-cond}
{\lambda}_{\mu|\nu} = 0, \ \ {\lambda}^{\mu}\,_{|\nu} = 0,
\end{equation}
where the comma denotes partial differentiation with
respect to the coordinate functions $x^\nu$ and the stroke denotes covariant differentiation with respect to
the connection $\Gamma^{\alpha}_{~\mu\nu}$. The connection (\ref{canonical}) will be referred to as the canonical (or Weitzenb\"{o}ck) connection. The relation (\ref{AP-cond}) is known in the literature as the AP-condition.
Let
\begin{equation}\label{torsion}
\Lambda^{\alpha}_{~\mu\nu}: = \Gamma^{\alpha}_{~\mu\nu} -
\Gamma^{\alpha}_{~\nu\mu}
\end{equation}
denote the torsion tensor of the canonical connection
$\Gamma^{\alpha}_{~\mu\nu}$.

The AP-condition (\ref{AP-cond}) together with the
commutation formula
$$\undersym{\lambda}{i}^{\alpha} \,_{|\mu\nu} - \undersym{\lambda}{i}^{\alpha} \,_{|\nu\mu} =
\undersym{\lambda}{i}^{\epsilon}P^{\alpha}_{~~\epsilon\mu\nu} + \undersym{\lambda}{i}^{\alpha}
\,_{|\epsilon}\, \Lambda^{\epsilon}_{~\nu\mu}$$
forces the curvature
tensor $P^{\alpha}_{~~\mu\nu\sigma}$ of the canonical connection
$\Gamma^{\alpha}_{~\mu\nu}$ to vanish identically \cite{NA07}.
However, an AP-space still have other connections which are non-flat, namely, the dual connection
\begin{equation}\label{dual}
\widetilde{{\Gamma}}^{\alpha}_{~\mu\nu}: = \Gamma^{\alpha}_{~\nu\mu},
\end{equation}
and the symmetric connection
\begin{equation}\label{symmetric}
\widehat{{\Gamma}}^{\alpha}_{~\mu\nu}: =
\frac{1}{2}(\Gamma^{\alpha}_{~\mu\nu} + \Gamma^{\alpha}_{~\nu\mu}) =
\Gamma^{\alpha}_{~(\mu\nu)}.
\end{equation}
Moreover, the parallelization vector fields define a metric tensor on $M$ by
\begin{equation}\label{metric}
g_{\mu \nu} := \e{\lambda}{i}{_{\mu}}\e{\lambda}{i}{_{\nu}}
\end{equation}
with inverse metric
\begin{equation}\label{inverse}
g^{\mu \nu} = \e{\lambda}{i}^{\mu}\e{\lambda}{i}^{\nu}.
\end{equation}
In view of (\ref{AP-cond}), the canonical connection $\Gamma^\alpha_{~\mu\nu}$ (\ref{canonical}) is metric: $g_{{\mu\nu}|\sigma}=0$\, and,\, consequently,\,  ${g^{\mu\nu}}_{|\sigma}=0$. The Levi-Civita connection associated with $g_{\mu\nu}$ is
\begin{equation}\label{Christoffel}
\overcirc{{\Gamma}}^{\alpha}_{~\mu\nu}= \frac{1}{2}g^{\alpha \sigma}\left(g_{\mu \sigma, \nu}+g_{\nu \sigma,\mu}-g_{\mu \nu,\sigma}\right).
\end{equation}
The Ricci tensor can be written as
\begin{equation}\label{Ricci}
    R_{\mu \nu}:=R^{\alpha}{_{\mu \nu \alpha}}=\overcirc{{\Gamma}}^{\alpha}_{~\mu\nu,\alpha}-\overcirc{{\Gamma}}^{\alpha}_{~\alpha \mu,\nu}
    +\overcirc{{\Gamma}}^{\alpha}_{~\mu\nu}~\overcirc{{\Gamma}}^{\beta}_{~\alpha\beta}
    -\overcirc{{\Gamma}}^{\beta}_{~\alpha \mu}~\overcirc{{\Gamma}}^{\alpha}_{~\beta \nu}.
\end{equation}
Summing up, an AP-space possesses at least four \textbf{natural}\footnote{By "a natural geometric object" we mean that this object is formed of the building blocks of the geometry \textbf{only}, i.e., in terms of the parallelization vector fields $\e{\lambda}{i}$ \textbf{only}.} linear connections: (\ref{canonical}), (\ref{dual}), (\ref{symmetric}) and (\ref{Christoffel}). The properties of these connections are summarized in Table \ref{T1}.

The contortion tensor $\gamma^{\alpha}_{~\mu\nu}$ is defined by any one of the following two equivalent \ formulae:
\begin{eqnarray}
\gamma^{\alpha}_{~\mu\nu} &=& \Gamma^{\alpha}_{~\mu\nu} - \overcirc{\Gamma}^{\alpha}_{~\mu\nu}, \label{contortion1}\\
  \gamma^{\alpha}_{~\mu\nu}  &=& \undersym{\lambda}{i}^{\alpha}\,\,
\undersym{\lambda}{i}_{\mu\,;\,\nu}, \label{contortion2}
\end{eqnarray}
where the semicolon denotes covariant differentiation with respect to \,$\overcirc{\Gamma}^{\alpha}_{~\mu\nu}$.
Since \ $\overcirc{\Gamma}^{\alpha}_{~\mu\nu}$ is symmetric, it
follows that (using (\ref{contortion1}))
\begin{equation}\label{torsion-contortion}
\Lambda^{\alpha}_{~\mu\nu} = \gamma^{\alpha}_{~\mu\nu} - \gamma^{\alpha}_{~\nu\mu}.
\end{equation}
One can also show that:
\begin{equation}\label{torsion03}
        \Lambda_{\alpha \mu \nu}=\gamma_{\alpha \mu \nu}-\gamma_{\alpha \nu \mu},
\end{equation}
\begin{equation}\label{contortion03}
        \gamma_{\alpha \mu \nu}=\frac{1}{2}\left(\Lambda_{\nu\alpha\mu}+\Lambda_{\alpha\mu\nu}-\Lambda_{\mu\alpha\nu}\right),
\end{equation}
where $\Lambda_{\mu\nu\sigma} =
g_{\epsilon\mu}\,\Lambda^{\epsilon}_{~\nu\sigma}$\, and \,$\gamma_{\mu\nu\sigma} =
g_{\epsilon\mu}\,\gamma^{\epsilon}_{~\nu\sigma}$. It is to be noted
that $\Lambda_{\mu\nu\sigma}$ is skew-symmetric in the last pair of
indices whereas $\gamma_{\mu\nu\sigma}$ is skew-symmetric in the
first pair of indices. Moreover, it follows, from (\ref{torsion03}) and (\ref{contortion03}), that the torsion
tensor vanishes if and only if the contortion tensor vanishes.
The contraction of the torsion or the contortion tensors produces a 1-form $C_\mu$, called the basic form:
\begin{equation}\label{basic-form}
        C_{\mu} := \Lambda^{\alpha}_{~\mu \alpha}=\gamma^{\alpha}_{~\mu \alpha}.
\end{equation}

\vspace{8pt}
It has been shown in \cite{W2000} that parameterizing the above mentioned four linear connections gives the following object
\begin{equation*}
\nabla^\alpha_{~\mu\nu}:=a_1 \Gamma^\alpha_{~\mu\nu}+ a_2 \widetilde{\Gamma}^\alpha_{~\mu\nu}+ a_3 \widehat{\Gamma}^\alpha_{~\mu\nu}
+a_4~ \overcirc{\Gamma}^\alpha_{~\mu\nu},
\end{equation*}
Assuming that this object is a linear connection, it can be written, after some calculations, as
\begin{equation}\label{p-canonical}
    \nabla^{\alpha}_{~\mu \nu}=\overcirc{\Gamma}^\alpha_{~\mu\nu}+b\, \gamma^{\alpha}_{~\mu \nu},
\end{equation}
where $b$ is a parameter to be discussed later. Geometrical and physical arguments for such parametrization are given in \cite{W2000}.

The linear connection (\ref{p-canonical}) will be referred to as the parameterized canonical connection. This connection is metric. It is also non-symmetric due to the non-symmetry of the contortion $\gamma^\alpha_{~\mu\nu}$. Consequently, its dual $\widetilde{\nabla}^\alpha_{~\mu\nu}$  and its symmetric part $\widehat{\nabla}^\alpha_{~\mu\nu}$ exist and are defined by:
 \begin{eqnarray}
   \widetilde{\nabla}^\alpha_{~\mu\nu}  &=&  \nabla^\alpha_{~\nu\mu}\label{p-dual}= \,\,\overcirc{\Gamma}^\alpha_{~\nu\mu}+b\, \gamma^{\alpha}_{~\nu \mu}, \\
   \widehat{\nabla}^\alpha_{~\mu\nu}    &=&  \frac{1}{2}\,(\nabla^\alpha_{~\mu\nu}+ \nabla^\alpha_{~\nu\mu}).\label{p-symm}
 \end{eqnarray}
These connections will be called parameterized dual and parameterized symmetric connections respectively. We then have, in addition to the four connections of the AP-space, three more connections: (\ref{p-canonical}), (\ref{p-dual}), and (\ref{p-symm}).

The parameterized torsion, parameterized contortion and parameterized basic form are then given respectively by:
  \begin{equation}\label{p-objects}
    \s{\Lambda}{^{\alpha}}_{\mu \nu}=b\,\Lambda{^{\alpha}}_{\mu \nu}, \quad  \s{\gamma}{^{\alpha}}_{\mu \nu}=b\, \gamma^{\alpha}_{~\mu \nu}, \quad \s{C}_{\mu}=b\,C_{\mu}.
  \end{equation}

When we consider the parameterized connections we will speak of parameterized AP-space (PAP-space) and parameterized AP-geometry (PAP-geometry). From now on, any geometric object living in a PAP-space will be decorated by an asterisk, meaning that it contains powers of the parameter $b$, starting from the first order. Consequently, the vanishing of $b$ will lead to the vanishing of this object.\\

Table \ref{T1} gives a summary of the connections and their associated geometric objects in AP-space \cite{NA07} and PAP-space. This table also clarifies the  discussion that follows (at the end of the present section).

\begin{table}
\begin{center}
\caption{Linear connections in AP-space and PAP-space}\label{T1}
\begin{tabular}{|c|c|c|c|c|c|c|}
\hline
                  &\multirow{2}{*}{connection} & connection  & covariant  & torsion & curvature & metricity \\
                  &        & coefficient & derivative & tensor  & tensor    &  \\
                  \hline
 \parbox[t]{2mm}{\multirow{8}{*}{\rotatebox[origin=c]{90}{AP-Space}}}
 &\multirow{2}{*}{canonical} & \multirow{2}{*}{$\Gamma^{\alpha}{_{\mu \nu}}$} & \multirow{2}{*}{${|}$} & \multirow{2}{*}{$\Lambda^{\alpha}{_{\mu \nu}}$} & \multirow{2}{*}{$0$} & \multirow{2}{*}{$\checkmark$} \\
                  &      &                                       &                                         & & & \\   \cline{2-7}
                  & \multirow{2}{*}{dual} & \multirow{2}{*}{$\widetilde{\Gamma}^{\alpha}{_{\mu \nu}}$} & \multirow{2}{*}{$\widetilde{|}$} & \multirow{2}{*}{$-\Lambda^{\alpha}{_{\mu \nu}}$} & \multirow{2}{*}{$\widetilde{R}^{\alpha}{_{\mu \nu \sigma}}$} & \multirow{2}{*}{$\times$} \\
                  &      &                                       &                                         & & & \\   \cline{2-7}
                  & \multirow{2}{*}{symmetric} & \multirow{2}{*}{$\widehat{\Gamma}^{\alpha}{_{\mu \nu}}$} & \multirow{2}{*}{$\widehat{|}$} & \multirow{2}{*}{$0$} & \multirow{2}{*}{$\widehat{R}^{\alpha}{_{\mu \nu \sigma}}$} & \multirow{2}{*}{$\times$} \\
                  &      &                                       &                                         & & & \\   \cline{2-7}
                  & Riemannian & \multirow{2}{*}{${\overcirc{\Gamma}}{^{\alpha}}{_{\mu \nu}}$} & \multirow{2}{*}{$;$} & \multirow{2}{*}{$0 $} &\multirow{2}{*}{${R}{^{\alpha}}{_{\mu \nu \sigma}}$}  & \multirow{2}{*}{$\checkmark$} \\
                  & (Levi-Civita) &  &  &  &  &  \\
                  \hline
 \parbox[t]{2mm}{\multirow{6}{*}{\rotatebox[origin=c]{90}{PAP-Space}}}
                  & parameterized & \multirow{2}{*}{$\nabla^{\alpha}{_{\mu \nu}} $} & \multirow{2}{*}{$\|$} & \multirow{2}{*}{$b~\Lambda^{\alpha}{_{\mu \nu}}$} & \multirow{2}{*}{$ B^{\alpha}{_{\mu \nu \sigma}}$} & \multirow{2}{*}{$\checkmark$} \\
                  & canonical &  &  &  &  &  \\ \cline{2-7}
                  & parameterized & \multirow{2}{*}{$ \widetilde{\nabla}^{\alpha}{_{\mu \nu}}$} & \multirow{2}{*}{$\widetilde{\|}$} & \multirow{2}{*}{$-b~\Lambda^{\alpha}{_{\mu \nu}}$} & \multirow{2}{*}{$\widetilde{B}^{\alpha}{_{\mu \nu \sigma}} $} & \multirow{2}{*}{$\times$} \\
                  & dual &  &  &  &  &  \\ \cline{2-7}
                  & parameterized & \multirow{2}{*}{$\widehat{\nabla}^{\alpha}{_{\mu \nu}} $} & \multirow{2}{*}{$\widehat{\|}$} & \multirow{2}{*}{$0 $} & \multirow{2}{*}{$\widehat{B}^{\alpha}{_{\mu \nu \sigma}} $} &\multirow{2}{*}{$\times$} \\
                  & symmetric &  &  &  &  &\\
                  \hline
\end{tabular}
\end{center}
\end{table}

Table \ref{T2} represents the PAP-version of Mikhail's table \cite{M62}. The later can be recovered by setting $b=1$, where the $||$ covariant derivative reduces to the $|$ one. The tensors mentioned in Table \ref{T2} are expected to be of extreme importance in physical applications as in the AP-case (cf., \cite{SO2010, MWE95, W85, W2007}). In Table \ref{T2}, the tensor $\s{\Delta}{^\alpha}{_{\mu \nu}}$ is defined by $\s{\Delta}{^\alpha}{_{\mu \nu}}=\s{\gamma}{^\alpha}{_{\mu \nu}}+ \s{\gamma}{^\alpha}{_{\nu \mu}}$. Since the AP-space is recovered, we may consider the following relation for the Ricci tensor (\ref{Ricci})
$${R}_{\mu \nu}=\frac{1}{2}(\psi_{\mu\nu}-\phi_{\mu\nu}-\theta_{\mu\nu})+\omega_{\mu\nu},$$
and the important identity
\begin{equation}\label{identity}
\eta_{\mu \nu}+\varepsilon_{\mu \nu}-\chi_{\mu \nu}\equiv 0.
\end{equation}
\begin{table}
\begin{center}
\caption{Second order world tensors in PAP-geometry}\label{T2}
\begin{tabular}{|c|c|}
\hline
  skew-symmetric& symmetric\\
\hline
  $\s{\xi}_{\mu \nu}:=\s{\gamma}_{\mu \nu}{^{\alpha}}_{\|{\alpha}}=b {\gamma}_{\mu \nu}{^{\alpha}}_{\|{\alpha}}$&  \\[5 pt]
  $\s{\zeta}_{\mu \nu}:=\s{C}_{\alpha}\s{\gamma}_{\mu \nu}{^{\alpha}}=b^2 {C}_{\alpha} {\gamma}_{\mu \nu}{^{\alpha}}$&  \\[3 pt]
\hline
 $\s{\eta}_{\mu \nu}:=\s{C}_{\alpha}\s{\Lambda}{^{\alpha}}_{\mu \nu}=b^2 {C}_{\alpha} {\Lambda}{^{\alpha}}_{\mu \nu}$ &
 $\s{\phi}_{\mu \nu}:=\s{C}_{\alpha}\s{\Delta}{^{\alpha}}_{\mu \nu}=b^2 {C}_{\alpha} {\Delta}{^{\alpha}}_{\mu \nu}$ \\[5 pt]
 $\s{\chi}_{\mu \nu}:=\s{\Lambda}{^{\alpha}}_{\mu \nu \| {\alpha}}=b {\Lambda}{^{\alpha}}_{\mu \nu \| {\alpha}}$ &
 $\s{\psi}_{\mu \nu}:=\s{\Delta}{^{\alpha}}_{\mu \nu \| {\alpha}}=b {\Delta}{^{\alpha}}_{\mu \nu \| {\alpha}}$  \\[3 pt]
 $\s{\varepsilon}_{\mu \nu}:=\s{C}_{{\mu}\|\nu}-\s{C}_{{\nu}\|\mu}=b ({C}_{{\mu}\|\nu}-{C}_{{\nu}\|\mu})$ &
 $\s{\theta}_{\mu \nu}:=\s{C}_{{\mu}\|\nu}+\s{C}_{{\nu}\|\mu}=b ({C}_{{\mu}\|\nu}+{C}_{{\nu}\|\mu})$  \\[3 pt]
 $\s{\kappa}_{\mu \nu}:=\s{\gamma}{^{\alpha}}_{\mu \sigma}\s{\gamma}{^{\sigma}}_{\alpha \nu}-\s{\gamma}{^{\alpha}}_{\nu \sigma}\s{\gamma}{^{\sigma}}_{\alpha \mu}$ &
 $\s{\varpi}_{\mu \nu}:=\s{\gamma}{^{\alpha}}_{\mu \sigma}\s{\gamma}{^{\sigma}}_{\alpha \nu}+\s{\gamma}{^{\alpha}}_{\nu \sigma}\s{\gamma}{^{\sigma}}_{\alpha \mu}$ \\[3 pt]
 \hspace{1 cm}$=b^2 ({\gamma}{^{\alpha}}_{\mu \sigma} {\gamma}{^{\sigma}}_{\alpha \nu}- {\gamma}{^{\alpha}}_{\nu \sigma} {\gamma}{^{\sigma}}_{\alpha \mu})$&  \hspace{1 cm}$=b^2 ({\gamma}{^{\alpha}}_{\mu \sigma} {\gamma}{^{\sigma}}_{\alpha \nu}+ {\gamma}{^{\alpha}}_{\nu \sigma} {\gamma}{^{\sigma}}_{\alpha \mu})$\\[3pt]
\hline
 $$ & $\s{\sigma}_{\mu \nu}:=\s{\gamma}{^{\alpha}}_{\sigma \mu} \s{\gamma}{^{\sigma}}_{\alpha \nu}=b^2 {\gamma}{^{\alpha}}_{\sigma \mu} {\gamma}{^{\sigma}}_{\alpha \nu}$ \\[5 pt]
 $$ & $\s{\omega}_{\mu \nu}:=\s{\gamma}{^{\alpha}}_{\mu \sigma} \s{\gamma}{^{\sigma}}_{\nu \alpha}=b^2 {\gamma}{^{\alpha}}_{\mu \sigma} {\gamma}{^{\sigma}}_{\nu \alpha}$  \\[5 pt]
 $$ & $\s{\alpha}_{\mu \nu}:=\s{C}_{\mu} \s{C}_{\nu}=b^2 {C}_{\mu} {C}_{\nu}$ \\[5 pt]
 $$ & \\
\hline
\end{tabular}
\end{center}
\end{table}

The main advantages of the parametrization process are:
\begin{enumerate}
  \item The most important connections in AP-geometry, namely, the canonical connection (\ref{canonical}) and the Levi-Civita connection (\ref{Christoffel}) are related via (\ref{p-canonical}) (together with (\ref{contortion1})).
  \item The parameterized canonical connection $\nabla^\alpha_{~\mu\nu}$ is the only \textbf{\emph{natural}} connection in the PAP-space which is \textbf{\emph{metric and having non-vanishing torsion and non-vanishing curvature}}. It is more interesting than the canonical connection which has no curvature and the Levi-Civita connection which has no torsion.
  \item The connection $\nabla^\alpha_{~\mu\nu}$ reduces to the canonical connection when $b=1$ and to the Levi-Civita connection when $b=0$.
  \item The parameter $``b"$ in (\ref{p-canonical}) plays an important role in physical applications, especially in geometric field theories \cite{SM04,G2007, W2012, WMK2000, WMK2001}. This point will be discussed later in Section \ref{S7}.
\end{enumerate}
\section{A Lagrangian Function in PAP-Geometry}\label{S3}
It is well known that the GR-theory can be formulated by choosing the Ricci scalar to construct a Lagrangian density in the Einstein-Hilbert action. This led some authors \cite{W75, W2001, WS2010} to examine the curvature scalars of the connections mentioned in Table \ref{T1}. Among these connections, we are going to consider the curvature scalar of the parameterized dual connection $\widetilde{\nabla}^\alpha_{~\mu\nu}$ to construct a new Lagrangian density. Our choice is motivated by the success of the Generalized Field Theory (GFT)
 (\cite{MW77}, \cite{MW81}) based on the dual connection $\widetilde{\Gamma}^\alpha_{\,\mu\nu}$ of the AP-geometry.
We next use a variational method, Dolan-McCrea method, to find out the field equations.\\

The curvature tensor $\widetilde{B}^\alpha_{~\mu\nu\sigma}$ of\, $\widetilde{\nabla}^\alpha_{~\mu\nu}$, calculated using (\ref{p-dual}), gives
\begin{equation}\label{b-dual curvature}
    \widetilde{B}^{\alpha}{_{\mu \nu \sigma}}=R^{\alpha}{_{\mu \nu \sigma}}+ \s{Q}{^{\alpha}}{_{\mu \nu \sigma}},
\end{equation}
where $R^{\alpha}_{~\mu \nu \sigma}$ is the Riemannian curvature tensor associated with (\ref{Christoffel}) and $\s{Q}$ is the tensor field of type (1,3) given by

\begin{equation}\label{Qstar}
    \s{Q}{^{\alpha}}{_{\mu \nu \sigma}}=\s{\gamma}{^{\alpha}}{_{\sigma \mu \,;\,\nu}}-\s{\gamma}{^{\alpha}}{_{\nu \mu \,;\, {\sigma}}}+\s{\gamma}{^{\epsilon}}_{\sigma \mu}\s{\gamma}{^{\alpha}}{_{\nu \epsilon}}-\s{\gamma}{^{\epsilon}}{_{\nu \mu}}\s{\gamma}{^{\alpha}}{_{\sigma \epsilon}},
\end{equation}
which is a parameterized tensor of type $(1,3)$ expressed in terms of the parameterized contortion (or torsion). Contracting the curvature tensor $\widetilde{B}{^{\alpha}}{_{\mu \nu
\sigma}}$, by setting $\alpha=\sigma$, we get
\begin{equation*}
    \widetilde{B}_{\mu \nu}=R_{\mu \nu}-\s{\gamma}{^\alpha}{_{\nu \mu \,;\, \alpha}}+\s{\gamma}{^{\epsilon}}{_{\alpha \mu}}\s{\gamma}{^{\alpha}}{_{\nu \epsilon}},
\end{equation*}
where $\widetilde{B}_{\mu \nu}:=\widetilde{B}^\alpha_{~\mu \nu\alpha}$ and $R_{\mu\nu}$ is the Ricci tensor.\\
Contracting again by $g^{\mu \nu}$, a scalar $\widetilde{B}:=g^{\mu \nu} \widetilde{B}_{\mu \nu}$ is given by
\begin{equation*}
    \widetilde{B}=R+\s{C}{^{{\alpha}}}{_{;\,\alpha}}+\frac{1}{2}{\s{\Lambda}{^{\epsilon}}}_{\alpha \mu} {\s{\gamma}{^{\alpha \mu}}}_{\epsilon}.
\end{equation*}

Now, let us define a Lagrangian density $\mathcal{L}$ by
\begin{equation}\label{q33}
    \mathcal{L} =\lambda \widetilde{B}= \lambda\,(R+\s{C}{^{{\alpha}}}{_{;\,\alpha}}+\frac{1}{2}{\s{\Lambda}{^{\epsilon}}}_{\alpha \mu} {\s{\gamma}{^{\alpha \mu}}}_{\epsilon}).
\end{equation}
It can be shown that $\lambda \s{C}{^{\alpha}}{_{;\,\alpha}}=(\lambda \s{C}{^{\alpha}})_{,\alpha}$. As this term has no contribution, the Lagrangian density takes the form
\begin{equation}\label{q34}
    \mathcal{L}=\lambda(R+\frac{1}{2}\,{\s{\Lambda}{^{\epsilon \alpha \mu}}}\, {\s{\gamma}_{\alpha\mu\epsilon}}).
\end{equation}
The density function appears as the Ricci scalar in addition to an extra scalar term that consists mainly of torsion.

We are going now to apply a variational method alternative to the action principle procedure. It has been suggested by Dolan and McCrea \cite{DM} in the context of Riemannian geometry. Dolan-McCrea's (DM-) method has been applied to AP-geometry, by Mikhail and Wanas, to construct a unified field theory \cite{MW77}. This method requires a scalar Lagrangian density function of the form
\begin{equation}\label{q35}
    \mathcal{L} = \mathcal{L}(\e{\lambda}{i}{_{\mu}},\e{\lambda}{i}{_{\mu,\nu}}, \e{\lambda}{i}{_{\mu,\nu \sigma}})=\lambda L,
\end{equation}
where $L(\e{\lambda}{i}{_{\mu}},\e{\lambda}{i}{_{\mu,\nu}}, \e{\lambda}{i}{_{\mu,\nu \sigma}})$ is a scalar field, function of $\e{\lambda}{i}{_{\mu}}$ and its first and second derivatives. One of the benefits of using DM-method is investigating differential identities admitted by a certain geometry. This allows studying conservation and choosing field equations. When applying the DM-method in AP-geometry we get the identity
\begin{equation}\label{DM_identity}
    H^{{\beta}}{_{{\nu}\,\widetilde{|}\,\beta}} \equiv 0,
\end{equation}
where
\begin{equation}\label{Hmn_def}
    H^{\beta}{_{\nu}} := \frac{\delta L}{\delta\e{\lambda}{i}{_{\beta}}}\e{\lambda}{i}{_{\nu}},
\end{equation}
and
\begin{equation}\label{del_deriv_def}
    \frac{\delta L}{\delta \e{\lambda}{i}{_{\beta}}} := \frac{1}{\lambda}\left[\frac{\partial \mathcal{L}}{\partial {\mathop{\lambda}\limits_{i}}{_\beta}}- \frac{\partial}{\partial x^{\gamma}}\left(\frac{\partial \mathcal{L}}{\partial {\mathop{\lambda}\limits_{i}}{_{\beta, \gamma}}}\right)+\frac{\partial^2}{\partial x^{\gamma} \partial x^{\sigma}}\left(\frac{\partial \mathcal{L}}{\partial {\mathop{\lambda}\limits_{i}}{_{\beta, \gamma \sigma}}}\right)\right].
\end{equation}
Considering the identity (\ref{DM_identity}) as a generalization of Bianchi second identity in the AP-space (or PAP, see theorems 1 and 2 of Ref. \cite{Wanas:2016boh}), we can assume that it gives, physically, a tensorial form of a type of conservation. Consequently, it gives rise to the field equations\footnote{As AP-geometry and PAP-geometry are defined by the same geometric structure $\e{\lambda}{i}$ and since the DM-method depends on a Lagrangian which is a function of $\e{\lambda}{i}$, the identity (\ref{DM_identity}) which have been obtained in the AP-geometry framework \cite{MW77} is still valid in PAP-geometry.}
\begin{equation}\label{q39}
    H_{\mu \nu}=0,
\end{equation}
where $H_{\mu \nu}$ is non-symmetric as clear from its definition (\ref{Hmn_def}), defined in the PAP geometry. Thus we have 16 field equations in 16 field variables $\e{\lambda}{i}{_{\mu}}$. As stated above the identity (\ref{DM_identity}) has been considered as a generalization of conservation. The conserved quantities are given by the components of the tensor $H_{\mu \nu}$.

\section{Field Equations}\label{S4}
In this Section we apply the DM-method to the Lagrangian density (\ref{q34}). We have \cite{W75}
\begin{equation*}
    \frac{\delta \lambda}{\delta \e{\lambda}{i}{_{\beta}}}=\frac{\partial \lambda}{\partial \e{\lambda}{i}{_{\beta}}}=\lambda~ \e{\lambda}{i}{^\beta}.
\end{equation*}
The $\delta$-derivation of the torsion tensor is
\begin{equation*}
    \frac{\delta \s{\Lambda}{^{\epsilon \alpha \mu}}}{\delta \e{\lambda}{i}{_{\beta}}}:=\frac{\partial \s{\Lambda}{^{\epsilon \alpha \mu}}}{\partial \e{\lambda}{i}{_{\beta}}}
    -\left(\frac{\partial \s{\Lambda}{^{\epsilon \alpha \mu}}}{\partial \e{\lambda}{i}{_{\beta,\gamma}}}\right)_{,\gamma},
\end{equation*}
where
\begin{eqnarray*}
  \frac{\partial \s{\Lambda}{^{\epsilon \alpha \mu}}}{\partial \e{\lambda}{i}{_{\beta}}} &=& -\e{\lambda}{i}{_\sigma}
  (g^{\beta \alpha}\s{\Lambda}{^{\epsilon \sigma \mu}}+g^{\beta \mu}\s{\Lambda}{^{\epsilon \alpha \sigma}})
  +\e{\lambda}{i}{^\epsilon} \s{\Lambda}{^{\beta \mu \alpha}}-\e{\lambda}{i}{^\alpha} \s{\Lambda}{^{\epsilon \beta \mu}}
  -\e{\lambda}{i}{^\mu} \s{\Lambda}{^{\epsilon \alpha \beta}}, \\
  \frac{\partial \s{\Lambda}{^{\epsilon \alpha \mu}}}{\partial \e{\lambda}{i}{_{\beta,\gamma}}} &=& g^{\beta \alpha} g^{\gamma \mu}
  \e{\lambda}{i}{^\epsilon}-g^{\gamma \alpha} g^{\beta \mu} \e{\lambda}{i}{^\epsilon}.
\end{eqnarray*}
Also, the $\delta$-derivation of the contortion tensor is
\begin{equation*}
    \frac{\delta \s{\gamma}{_{\alpha \mu \epsilon}}}{\delta \e{\lambda}{i}{_{\beta}}}:=\frac{\partial \s{\gamma}{_{\alpha \mu \epsilon}}}
    {\partial \e{\lambda}{i}{_{\beta}}}
    -\left(\frac{\partial \s{\gamma}{_{\alpha \mu \epsilon}}}{\partial \e{\lambda}{i}{_{\beta,\gamma}}}\right)_{,\gamma},
\end{equation*}
where
\begin{eqnarray*}
  \frac{\partial \s{\gamma}{_{\alpha \mu \epsilon}}}{\partial \e{\lambda}{i}{_{\beta}}} &=& \frac{1}{2}\e{\lambda}{i}{_\sigma}
  (\delta^{\beta}_{\alpha}\s{\Lambda}{^{\sigma}}{_{\mu \epsilon}}+\delta^{\beta}_{\mu}\s{\Lambda}{^{\sigma}}{_{\epsilon \alpha}}
  -\delta^{\beta}_{\epsilon}\s{\Lambda}{^{\sigma}}{_{\alpha \mu}}),\\
  \frac{\partial \s{\gamma}{_{\alpha \mu \epsilon}}}{\partial \e{\lambda}{i}{_{\beta,\gamma}}} &=& \frac{1}{2}\e{\lambda}{i}{_{\mu}}\delta^{\beta \gamma}_{[\epsilon \alpha]}
  +\frac{1}{2}\e{\lambda}{i}{_{\epsilon}}\delta^{\beta \gamma}_{[\mu \alpha]}+\frac{1}{2}\e{\lambda}{i}{_{\alpha}}\delta^{\beta \gamma}_{[\mu \epsilon]},
\end{eqnarray*}
with $\delta^{\beta \gamma}_{[\epsilon \alpha]}:=\delta^{\beta}_{\epsilon} \delta^{\gamma}_{\alpha}-
\delta^{\beta}_{\alpha} \delta^{\gamma}_{\epsilon}$.
Substituting the above differential quantities into (\ref{Hmn_def}), using (\ref{del_deriv_def}), the field equations (\ref{q39}) read
\begin{equation}\begin{split}\label{field-eqns}
    H_{\mu \nu}=&-2G_{\mu \nu}+g_{\mu \nu} \s{\gamma}{^{\epsilon \alpha \sigma}}\,\s{\gamma}_{\alpha \sigma \epsilon}-2\s{\gamma}{^{\alpha}}_{\mu \epsilon} \s{\gamma}{^{\epsilon}}_{\alpha \nu}\\+&2b(\s{\gamma}{^\alpha}_{\mu \nu \| {\alpha}}-\s{C}{_{\alpha}} \s{\gamma}{^{\alpha}}_{\mu \nu}+\s{\gamma}{^{\alpha}}_{\mu \epsilon} \s{\gamma}{^{\epsilon}}_{\alpha \nu}+\s{\gamma}{^{\alpha}}_{\mu \epsilon} \s{\gamma}{^{\epsilon}}_{\nu \alpha})=0,
\end{split}\end{equation}
where $G_{\mu\nu}$ is the Einstein tensor.
It is clear that $H_{\mu \nu}$ is a non-symmetric tensor defined in PAP-geometry.

In the next section we discuss the physical consequences of these field equations.


\section{Physical Analysis of the Theory}\label{S5}
In order to give some physical information about the theory suggested, we are going to apply three ideas. These are to be applied before solving the equations of the theory. The first (Subsection \ref{S5.1}) is comparing the symmetric and skew-symmetric parts of the field equations with nonlinear field theories, GR and Einstein-Maxwell theories. The second (Subsection \ref{S5.2}) is comparing the content of the theory with linear field theories, Newton and Maxwell's theories. The third (Subsection \ref{S5.3}) is examining the capability of the theory to represent physical fields and their strengths.

\subsection{Comparison with nonlinear field theories}\label{S5.1}
\paragraph{Symmetric part of the field equations.} Using the tensors of Table \ref{T2}, the symmetric part of the field equations (\ref{field-eqns}) is given by
\begin{equation}\label{symm-field-eqns}
H_{(\mu\nu)}=-2G_{\mu \nu}-(\s{\varpi}_{\mu \nu}-\frac{1}{2}g_{\mu \nu} \s{\varpi})+b (\s{\psi}_{\mu \nu}-\s{\phi}_{\mu \nu}+\s{\varpi}_{\mu \nu}+2\s{\omega}_{\mu \nu})=0.
\end{equation}
Consequently, it can be written as
\begin{equation}\label{q42}
    R_{\mu \nu}-\frac{1}{2}\,g_{\mu \nu} R = -\frac{1}{2}\,\s{T}_{\mu \nu},
\end{equation}
where
\begin{equation}\label{material-energy}
    \s{T}_{\mu \nu}:=(\s{\varpi}_{\mu \nu}-\frac{1}{2}g_{\mu \nu} \s{\varpi})-b (\s{\psi}_{\mu \nu}-\s{\phi}_{\mu \nu}+\s{\varpi}_{\mu \nu}+2\s{\omega}_{\mu \nu}).
\end{equation}
Comparing with the GR, we can take the LHS of the field equations (\ref{q42}) as representing the gravitational field and $g_{\mu \nu}$ as representing the gravitational potential. The RHS is suggested to  represent a material-energy distribution. As the LHS of (\ref{q42}) satisfies the contracted Bianchi differential identity, the RHS gives rise to the conservation relation
\begin{equation}\label{q44}
    \s{T}{^{\mu \nu}}{_{;\,\mu}}=0.
\end{equation}
One of the main differences between the field equations (\ref{q42}) of the present theory and the corresponding field equations of GR is that the material-energy distribution (\ref{material-energy}) is given as a pure geometric object while it has been given phenomenologically in GR. This is an advantage of the theory and will be discussed later.\\

\paragraph{Skew-symmetric part of the field equations.} Using the tensors of Table \ref{T2}, the skew-symmetric part of (\ref{field-eqns}) is given by
\begin{equation}\label{skew-field-eqns}
H_{[\mu \nu]}=
b(\s{\chi}_{\mu \nu}-\s{\eta}_{\mu \nu})+(b-1)\s{\kappa}_{\mu \nu}=0.
\end{equation}
One can show that $\s{\eta}_{\mu\nu}=b^{2}\eta_{\mu\nu}$, $\s{\kappa}_{\mu\nu}=b^{2}\kappa_{\mu\nu}$ and $\s{\chi}_{\mu\nu}=b\chi_{\mu\nu}+b(b-1)(\eta_{\mu\nu}-\kappa_{\mu\nu})$. After some manipulation, taking the identity (\ref{identity}) into account, the above field equations reduce to
\begin{equation}\label{q46}
b^{2}\varepsilon_{\mu\nu}=0,
\end{equation}
which can be written as
\begin{equation}
\nonumber   b(\s{C}_{\mu,\nu}-\s{C}_{\nu,\mu})=\s{\eta}_{\mu\nu}.
\end{equation}
So, if we set
\begin{equation}\label{q47}
    \s{F}_{\mu \nu}:=\s{\eta}_{\mu \nu},
\end{equation}
the skew part of the field equations is written as
\begin{equation}\label{electromagnetic}
    \s{F}_{\mu \nu}=b(\s{C}_{\mu, \nu}-\s{C}_{\nu, \mu}).
\end{equation}

Comparing with the Einstein-Maxwell theory, we suggest to take the basic form $\s{C}_{\mu}$ to represent the electromagnetic potential and (\ref{q47}) to describe the electromagnetic field strength. This will be made more clear in the next section.
\subsection{Comparison with linear field theories}\label{S5.2}

In classical linear field theories the equations are sometimes quantitatively poor but qualitatively clear. Contrarily, in modern theories the equations are highly nonlinear, which quantifies accumulating small effects. But the physical interpretation of the equations' structure becomes more difficult. This difficulty can sometimes be overcame by linearizing the equations when assuming a weak static field and slow motion. So, it is convenient to write the field equations (\ref{field-eqns}) in a linearized form in order to compare them with the classical linear field theories and to clarify the physical entities of the equations.

In the following we let $\dim\, M=4$. Then, we expand the tetrad fields $\e{\lambda}{i}$ ($i=1,\cdots 4$) as
\begin{equation}\label{q49}
    \e{\lambda}{i}{_\mu}=\delta_{i \mu}+\epsilon \e{h}{i}{_\mu},
\end{equation}
where $\delta_{i \mu}$ is the kroneckar delta, $\e{h}{i}{_\mu}$ are functions representing the deviation of the tetrad from being Galilean and $\epsilon$ is a small parameter such that we can neglect $\epsilon^2$ and higher orders. Now, we are able to expand any function of $\e{\lambda}{i}$ (e.g., a second order tensor components $A_{\mu \nu}$) as follows
\begin{equation*}
    A_{\mu \nu}=\n{A}_{\mu \nu}+\nn{A}_{\mu \nu}+\nnn{A}_{\mu \nu}+ ...,
\end{equation*}
where $\n{A}_{\mu \nu}$ is of zero order in $\epsilon$, $\nn{A}_{\mu \nu}$ is linear in $\epsilon$, $\nnn{A}_{\mu \nu}$ is quadratic in $\epsilon$, and so on. In what follows we neglect the quadratic and the higher terms in $\epsilon$.

Now, expanding $\e{\lambda}{i}^\mu$, $g_{\mu\nu}$ and $g^{\mu\nu}$, we get
\begin{eqnarray}
 \e{\lambda}{i}^{\mu} &=&\n{\e{\lambda}{i}}{^\mu}+\nn{\e{\lambda}{i}}{^\mu}+O(\epsilon^2)=\delta_{i\mu}-\epsilon~\e{h}{\mu}{_i}+O(\epsilon^2),\label{q50}\\
      {g}{_{\mu \nu}} &=& \n{g}{_{\mu \nu}}+\nn{g}{_{\mu \nu}}+O(\epsilon^2)=\delta_{\mu\nu}+\epsilon~ y_{\mu \nu}+O(\epsilon^2), \label{q51}\\
      {g}{^{\mu \nu}} &=& \n{g}{^{\mu \nu}}+\nn{g}{^{\mu \nu}}+O(\epsilon^2)=\delta_{\mu \nu}-\epsilon~ y_{\mu \nu}+O(\epsilon^2), \label{q52}
\end{eqnarray}
where
\begin{equation}\label{yy}
y_{\mu \nu}:=\e{h}{\mu}{_\nu}+\e{h}{\nu}{_\mu}.
\end{equation}
Moreover, the canonical connection (\ref{canonical}) and the Riemannian connection (\ref{Christoffel}) can be written respectively as
\begin{eqnarray}
    \nn{\Gamma}{^\alpha}{_{\mu \nu}} &=& \epsilon~ \e{h}{\alpha}{_{\mu,\nu}},\label{lin_can_conn}\\
    \nn{\overcirc{\Gamma}}{^{\alpha}}{_{\mu \nu}} &=& \frac{\epsilon}{2}\,(y_{\mu \alpha, \nu}+y_{\nu \alpha, \mu}- y_{\mu \nu, \alpha}).\label{lin_crs_symb}
\end{eqnarray}
Also, substituting from (\ref{lin_can_conn}) and (\ref{lin_crs_symb}) into (\ref{contortion1}) and using (\ref{torsion-contortion}), (\ref{basic-form}) and (\ref{p-objects}), we get the linearized expressions
\begin{equation}\label{q55}
\nn{\s{\Lambda}}{^\alpha}{_{\mu \nu}}= \epsilon (\e{h}{\alpha}{_{\mu,\nu}}-\e{h}{\alpha}{_{\nu,\mu}}),\,\,\,
\nn{\s{\gamma}}{^\alpha}{_{\mu \nu}}= \epsilon \e{h}{\alpha}{_{\mu, \nu}}-\frac{\epsilon}{2}(y_{\mu \alpha, \nu}+y_{\nu \alpha, \mu}- y_{\mu \nu, \alpha}),\,\,\,
\nn{\s{C}}{_\mu}= \epsilon (\e{h}{\alpha}{_{\mu,\alpha}}-\e{h}{\alpha}{_{\alpha,\mu}}).
\end{equation}
Similarly, all the fundamental tensors given in Table \ref{T2} can be linearized. For example,
\begin{equation}\label{q58}
\nn{\s{\Delta}}{^\alpha}{_{\mu \nu}}= \epsilon \,(y_{\mu \nu,\alpha}-\e{h}{\mu}{_{\alpha,\nu}}-\e{h}{\nu}{_{\alpha,\mu}}),
\end{equation}
Table \ref{T3} \cite{W75} classifies the fundamental tensors of Table \ref{T2} according to the order of $\epsilon$ appearing in their expanded expressions.
\begin{table}
\begin{center}
\caption{Expansion of fundamental tensors in terms of the order of $\epsilon$}\label{T3}
\begin{tabular}{|c|c|c|c|c|} \hline
\multirow{2}{*}{Geometric Objects} &Terms of &  Terms of & Terms of  & Terms of third\\
& 0-order &  first order & second order  & and higher order  \\
\hline $\h{i}{_\mu}$&$\checkmark$&$\checkmark$&$\times$&$\times$ \\
\hline
$g_{\mu\nu}$&$\checkmark$&$\checkmark$&$\checkmark$&$\times$  \\[3pt] \hline
$\h{i}{^\mu},~g^{\mu\nu}$&$\checkmark$&$\checkmark$&$\checkmark$&$\checkmark$ \\ \hline
$\overcirc{{\Gamma}}^{\alpha}_{~\mu\nu},~\s{\gamma}{^{\alpha}}_{\mu\nu},~\s{\nabla}{^{\alpha}}{_{\mu\nu}},$&\multirow{2}{*}{$\times$}&\multirow{2}{*}{$\checkmark$}&\multirow{2}{*}{$\checkmark$}& \multirow{2}{*}{$\checkmark$} \\[1 pt]
$\s{\Lambda}{^{\alpha}}{_{\mu\nu}},~\s{\Delta}{^{\alpha}}{_{\mu\nu}},~\s{C}_{\alpha}$&&&& \\ \hline
$\s{\xi}_{\mu\nu},~\s{\chi}_{\mu\nu},~\s{\varepsilon}_{\mu\nu}$&$\times$&$\checkmark$&$\checkmark$& $\checkmark$ \\ \hline
$\s{\theta}_{\mu\nu},~\s{\psi}_{\mu\nu},~R_{\mu \nu}$&$\times$&$\checkmark$&$\checkmark$& $\checkmark$ \\ \hline
$\s{\zeta}_{\mu\nu},~\s{\eta}_{\mu\nu},~\s{\kappa}_{\mu\nu}$&$\times$&$\times$&$\checkmark$& $\checkmark$ \\ \hline
$\s{\phi}_{\mu\nu},~\s{\varpi}_{\mu\nu},~\s{\omega}_{\mu\nu},$&\multirow{2}{*}{$\times$}&\multirow{2}{*}{$\times$}&\multirow{2}{*}{$\checkmark$}& \multirow{2}{*}{$\checkmark$} \\ [1 pt]
$\s{\sigma}_{\mu\nu},~\s{\alpha}_{\mu\nu}$&&&& \\
\hline
\end{tabular}
\end{center}
\end{table}

The main aim of this part of the work is to linearize the symmetric part (\ref{symm-field-eqns}) and the skew-symmetric part (\ref{skew-field-eqns}) of the field equations, which allows a comparison between the linearized symmetric equation and Newton's theory of gravity and a comparison between the linearized skew-symmetric equation and Maxwell's theory of electromagnetism.\\

\paragraph*{a. Linearizing the symmetric part of the field equations.} The symmetric part (\ref{q42}) of the field equations (\ref{field-eqns}) can be written in the alternative form:
\begin{equation}\label{symm-field-eqns_2}
    R_{\mu \nu}=-\frac{1}{2}(\s{T}{_{\mu \nu}}-\frac{1}{2}g_{\mu \nu} \s{T}),
\end{equation}
where $\s{T}:=g^{\mu\nu}\, \s{T}{_{\mu \nu}}$.
Using (\ref{material-energy}) together with Table \ref{T2}, the linearized form of equations (\ref{symm-field-eqns_2}) is written as
\begin{equation}\label{Lin-symm-FE}
    \nn{R}{_{\mu \nu}}=\frac{b}{2}(\nn{\s{\psi}}{_{\mu \nu}}-\frac{1}{2}\delta_{\mu \nu} \nn{\s{\psi}}),
\end{equation}
where $\s{\psi}:=g^{\mu\nu}\, \s{\psi}_{\mu \nu}$. However, the linearized Ricci tensor can be evaluated by inserting (\ref{lin_crs_symb}) into (\ref{Ricci}), which gives
\begin{equation}\label{Lin-RicciT}
    \nn{R}{_{\mu \nu}}=\frac{1}{2}(y_{\mu\nu,\alpha\alpha}-y_{\mu\alpha,\alpha\nu}-y_{\nu\alpha,\alpha\mu}+y_{\alpha\alpha,\mu\nu}).
\end{equation}
We first consider the case of vanishing $b$, which gives rise to a gravitational field in an empty space as clearly shown by (\ref{Lin-symm-FE}). Using the harmonic gauge conditions $\overcirc{\Gamma}{^\mu}{_{\alpha\alpha}}=0$, i.e. $y_{\mu\alpha,\alpha}=\frac{1}{2} y_{\alpha\alpha,\mu}$, the linearized form of the field equations (\ref{Lin-symm-FE}) in an empty space $R_{\mu\nu}=0$ can be written as\footnote{In (\ref{GW})-(\ref{GW-production}) we have relaxed the assumption of the static field.}
\begin{equation}\label{GW}
    \square^{2}~ y_{\mu\nu}=0,
\end{equation}
where $\square^{2}$ is the d'Alembert operator. The plane wave solution is a useful particular solution for the wave equation (\ref{GW}), which is given by
\begin{equation}\label{plane-wave}
    y_{\mu\nu}=\mathcal{C}_{\mu\nu}e^{i k_{\alpha}x^{\alpha}},
\end{equation}
where the symmetric tensor $\mathcal{C}_{\mu\nu}$ is the polarization tensor in the linearized regime and $k^{\alpha}$ is the wave vector. Since not all the components of $y_{\mu\nu}$ vanish everywhere, one can find that the plane wave (\ref{plane-wave}) is a solution to the linearized field equations (\ref{GW}) when the following two conditions are fulfilled \cite{Carroll:2004}: (i) the wave vector is null, i.e. $k_{\alpha}k^{\alpha}=0$; and (ii) the wave vector is orthogonal to $\mathcal{C}^{\mu\nu}$, i.e. $k_{\mu}\mathcal{C}^{\mu\nu}=0$. In this sense, one can interpret the transmission of the gravitation in free space as gravitational waves propagating at the speed of light. These waves can be fully characterized by $k^{\alpha}$ and $\mathcal{C}_{\mu\nu}$. The temporal component of the wave vector is the frequency of the wave, while the non vanishing components of $\mathcal{C}^{\mu\nu}$ give two modes of polarizations of a gravitational wave representing two states of a massless spin-$2$ particle (graviton).

We have shown that the propagation of gravitation in an empty space can be covered by the present theory in the same sense as in GR. However, the more general case $b\neq 0$ implies that the linearized form of the field equations (\ref{Lin-symm-FE}) to be
\begin{equation}\label{GW-production}
    \square^{2}~ y_{\mu\nu}=\frac{b}{2}(\nn{\s{\psi}}{_{\mu \nu}}-\frac{1}{2}\delta_{\mu \nu} \nn{\s{\psi}}),
\end{equation}
The above equation may have a novel feature characterizing the production of gravitational radiation by sources, where the non-minimal coupling between matter and gravity is adopted. This relates the tensors that describe the matter distribution (\ref{material-energy}) to some important physical quantities, especially when the higher order terms are accounted. To clarify this point we give an example by considering the case $\mu=\nu=4$ (the temporal component) in the above equation. This allows to write
\begin{equation}\label{RHS}
    \nabla^2\, y_{44}=b\left(\nn{\s{\psi}}{_{44}}-2(\nn{\s{\psi}}{_{11}}+\nn{\s{\psi}}{_{22}}
    +\nn{\s{\psi}}{_{33}}+\nn{\s{\psi}}{_{44}})\right),
\end{equation}
where $\nabla^{2}$  is the $3$-dimensional Laplacian. It seems that the tensor
$\s{\psi}{_{\mu \nu}}$ is responsible for describing the material distribution in its linearized form.

For an isotropic perfect fluid we may define
\begin{equation*}
    p_0:=\nn{\s{\psi}}{_{11}}\simeq\nn{\s{\psi}}{_{22}}\simeq\nn{\s{\psi}}{_{33}},\quad
    \rho_0:=\nn{\s{\psi}}{_{44}},
\end{equation*}
where $p_0$ and $\rho_0$ denote the proper pressure and proper density of the fluid (measured in geometric units), respectively.

So far we have used the weak and static filed assumption. Recalling the kinetic theory of gases, one can easily find out a quadratic behavior, in velocity, of the fluid pressure $p_0$. Accordingly, the spatial components of ${\s{\psi}}{_{\mu \nu}}$ are quadratic in velocity. Imposing the slowly moving test particle assumption, indeed, leads to neglecting the contribution of the spatial components of ${\s{\psi}}{_{\mu \nu}}$. Consequently, the linearized form of the symmetric field equations (\ref{RHS}) becomes
\begin{equation}\label{Lin-symm-FE2}
    \nabla^2 y_{44}=-b \rho_0.
\end{equation}

Now, let us define the function
\begin{equation}\label{Npotential}
    \phi:=k(g_{44}-1)=(k\epsilon) y_{44}=a y_{44},
\end{equation}
by (\ref{q51}), where $a$ is a parameter adjusting the order of magnitude of $\phi$. Then, equation (\ref{Lin-symm-FE2}) can be written as
\begin{equation}\label{poisson1}
    \nabla^2 \phi = - a b \rho_0,
\end{equation}
This equation has a form similar to the classical \textit{Poisson's equation}, which is
\begin{equation}\label{poisson2}
    \nabla^2 \Phi_N=-\frac{8 \pi G}{c^2}\rho,
\end{equation}
where $\Phi_N$ is the Newton gravitational potential, $G$ is the Newton gravitational constant and $c$ is the speed of light. Comparing the above two equations, (\ref{poisson1}) may be considered as \  Poisson's equation provided that we take $\rho=c^2 \rho_0$\, and\, $ab=8 \pi G$. This supports the following attributions: the gravitational potential to $\phi$\, or\, $g_{\mu \nu}$, via (\ref{Npotential}), and the material-energy distribution to $\s{T}_{\mu \nu}$.

On the other hand, if we take $\rho_0=0$ or $b=0$, then the field equations reduce to \textit{Laplace's equation}
\begin{equation}\label{Laplace}
    \nabla^2 \phi=0,
\end{equation}
which gives a further support to the above attributions.\\

\paragraph*{b. Linearizing the skew part of the field equations.} In order to compare the skew part of the suggested field equations with Maxwell's theory, we express (\ref{electromagnetic}) in its linearized form:
\begin{equation}\label{Lin-skew-FE}
    \nn{\s{F}}_{\mu \nu}=b(\nn{\s{C}}_{\mu, \nu}-\nn{\s{C}}_{\nu, \mu})=0.
\end{equation}
It is to be noted that the vanishing of $\nn{\s{F}}_{\mu \nu}$ in (\ref{Lin-skew-FE}) follows from (\ref{q47}) and Table \ref{T3}.
Making use of (\ref{q46}), (\ref{q47}) and Table \ref{T3}, one concludes that the electromagnetic field strength  $\s{F}_{\mu \nu}$ has no linear behavior. Consequently, the above equations do not represent conventional electromagnetic field that is induced by an electromagnetic potential $\s{C}_{\mu}$. In other words, the above equations will not reduce to the Maxwell's equations. This leads us to the conclusion that the field theory suggested is a pure gravity theory. This result will be justified in Section \ref{S6}.
\subsection{Type analysis}\label{S5.3}

The third tool is called $``$Type Analysis". It is not well known in the literature although it has been proved to be useful in the applications of any field theory constructed in AP-geometry \cite{SO2010, SO2011, W85, W2007}. Type analysis was originally suggested by Mikhail and Wanas \cite{MW81} in the context of the GFT \cite{MW77}. It gives the capability of any AP-structure to represent physical entities. The type analysis depends mainly on the expansion Table \ref{T3}, which is theory independent and can be used for both AP and PAP structures. Although the linearization scheme related to this table is not generally covariant, the type analysis is a covariant tool since it is based on vanishing of tensors.

For example, in the context of Riemannian geometry, a metric with vanishing curvature tensor is not appropriate to represent gravity and is characterized by the code $G0$, while a metric with non-vanishing curvature tensor can be used to represent gravity and is given the code $GI$.

In AP- (or PAP-) spaces, there are many tensors which enable to attribute more codes, corresponding to the tensor expressions constituting the field equations. In the context of the theory constructed in the present work, it is shown in subsection \ref{S5.2} that the theory is free from (conventional) electromagnetism. So, we manufacture the type codes for pure gravity. According to the order of $\epsilon$ appearing in the tensors constituting the symmetric part of the field equations, we write this symmetric part in the form
\begin{equation*}
    G_{\mu \nu}=-\frac{1}{2}\s{T}_{\mu \nu},
\end{equation*}
where, using (\ref{material-energy}),
\begin{equation*}
    \s{T}_{\mu \nu}=-b \s{\psi}_{\mu \nu}+\s{N}_{\mu \nu}-\frac{1}{2}g_{\mu \nu} \s{\varpi},
\end{equation*}
and
\begin{equation*}
\s{N}_{\mu \nu}:=
(1-2b)\s{\omega}_{\mu \nu} + b(\s{\phi}_{\mu \nu}-2\s{\omega}_{\mu \nu}).
\end{equation*}
The above tensors, together with the curvature tensor $R^{\alpha}{_{\mu \nu \sigma}}$ are those responsible for the type. Table \ref{T4} gives the different codes for gravity and material distribution.
\begin{table}
\begin{center}
 \caption{Type Analysis}\label{T4}
\begin{tabular}{|c|c|c|} \hline
Tensors  & Physical Meaning  & Code\\
\hline
$R^{\alpha}{_{\beta\gamma\delta}} = 0$, $\s{T}_{\mu \nu}=0$  & neither gravity nor material distribution & $G0$ \\[3pt]
\hline
$R^{\alpha}{_{\beta\gamma\delta}} \neq 0$, $\s{T}_{\mu\nu} = 0$ & gravitational field in free space & $GI$ \\ [5pt]
\hline
$R^{\alpha}{_{\beta\gamma\delta}} \neq 0$, $\s{\psi}_{\mu\nu} \neq 0$, $\s{N}_{\mu \nu}=0$, $\s{\varpi} = 0$ & gravitational field within  & $GII$ \\
&a material distribution & \\
\hline
$R^{\alpha}{_{\beta\gamma\delta}} \neq 0$, $\s{\psi}_{\mu\nu} \neq 0$, $\s{N}_{\mu\nu} \neq 0$, $\s{\varpi} = 0$ & strong gravitational field  & $GIII$ \\
&within a material distribution & \\
\hline
$R^{\alpha}{_{\beta\gamma\delta}} \neq 0$, $\s{\psi}_{\mu\nu} \neq 0$, $\s{N}_{\mu\nu} \neq 0$, $\s{\varpi} \neq 0$ & very strong gravitational field  & $GIV$ \\
& within a material distribution.& \\
\hline
\end{tabular}
\end{center}
\end{table}

\section{Solution in the Case of Spherical Symmetry}\label{S6}
In this section we aim to apply the field equations (\ref{field-eqns}) to the spherical symmetry case. This allows us to construct a model suitable to describe some astrophysical systems.\\
For the spherical symmetry case, the most general AP-structure constructed by Robertson \cite{R32} is given, in the coordinate system ($r$, $\theta$, $\varphi$, $t$), by
\begin{equation}\label{SS-Tetrad}
(\e{\lambda}{i}^{\mu})=\left(
  \begin{array}{cccc}
    B \sin{\theta} \cos{\varphi} & \frac{B}{r} \cos{\theta} \cos{\varphi} & -\frac{B \sin{\varphi}}{r \sin{\theta}} & 0 \\[5pt]
    B \cos{\theta} \cos{\varphi} & \frac{B}{r} \cos{\theta} \sin{\varphi} & \frac{B \cos{\varphi}}{r \sin{\theta}} & 0 \\[5pt]
    B \cos{\theta} & -\frac{B}{r} \sin{\theta} & 0 & 0 \\[5pt]
    rE & 0 & 0 & A \\
  \end{array}
\right),
\end{equation}
where $A$, $B$ and $E$ are functions of $r$ only.

In the context of the present field theory, evaluating the tensors listed in the first column of Table \ref{T4}, one can show that the tetrad (\ref{SS-Tetrad}) is of type $GIV$. This means that it is capable of representing strong gravitational field within a material distribution.

Using (\ref{metric}) and (\ref{SS-Tetrad}), the covariant components of the metric tensor are
\begin{equation}\label{cov_met}
(g_{\mu \nu})=\left(
  \begin{array}{cccc}
   \frac{1}{B^2} & 0 & 0 & -\frac{r E}{A B^2}\\[5pt]
   0 & \frac{r^2}{B^2} & 0 & 0\\[5pt]
   0 & 0 & \frac{r^2 \sin^2{\theta}}{B^2} & 0\\[5pt]
   -\frac{r E}{A B^2} & 0 & 0 & \frac{r^2 E^2+B^2}{A^2 B^2}
  \end{array}
\right),
\end{equation}

Making use of (\ref{SS-Tetrad}), (\ref{cov_met}) and the tensors of Table \ref{T2}, the field equations (\ref{field-eqns}) give rise to a highly nonlinear system of second order differential equations in $A(r)$, $B(r)$ and $E(r)$. A solution of one of these equations, consistent with the others, is $E(r)=0$.
Letting $E(r)=0$ in the above mentioned system, the field equations reduce to the following system:

\begin{equation}\label{FEb1}
\begin{split}
b^2\left[\frac{2B'^2}{B^2}-\frac{4B'}{rB}+\frac{A'^2}{A^2}\right]+\frac{2B'^2}{B^2}+\frac{4B'A'}{AB}-\frac{4B'}{rB}-\frac{4A'}{rA}=0,
\end{split}
\end{equation}

\begin{equation}\label{FEb2}
\begin{split}
&b^2 \left[\frac{2r^2B'^2}{B^2}-\frac{2r^2B''}{B}-\frac{r^2A'^2}{A^2}+\frac{2r^2B'A'}{AB}-\frac{2rB'}{B}\right]-\frac{2r^2B''}{B}+
\frac{4r^2A'^2}{A^2}\\
&-\frac{2r^2A''}{A}+\frac{2r^2B'^2}{B^2}-\frac{2rA'}{A}-\frac{2rB'}{B}=0,
\end{split}
\end{equation}

\begin{equation}\label{FEb3}
\begin{split}
b^2\left[\frac{-2B^2A''}{A^3}-\frac{2B'^2}{A^2}+\frac{2B'BA'}{A^3}+\frac{3B^2A'^2}{A^4}-\frac{4B^2A'}{rA^3}\right]-\frac{4BB''}{A^2}
+\frac{6B'^2}{A^2}-\frac{8B'B}{rA^2}=0.
\end{split}
\end{equation}

Examining the tensor responsible for the type, after the vanishing of the function $E(r)$, we find that the type is still $GIV$. This point will be discussed in the next section.

\vspace{10pt}
\noindent\emph{The exterior solution:}\vspace{3pt}

This case is interesting for various reasons. Firstly, it facilitates comparison with one of the most successful solutions of GR, the Schwarzschild exterior solution. Secondly, it provides an implicit test of the type analysis scheme. Thirdly, it helps explaining the capability of the suggested theory to cover known physics in this special case.

Using the definitions of the tensors in the first column of Table \ref{T4}, we can easily show that if $b=0$, the type of the tetrad (\ref{SS-Tetrad}) reduces to $GI$. This means that it is capable of representing a static gravitational field outside a spherically symmetric material distribution.
Now, the above system for $b=0$ reduces to
\begin{equation}\label{FE1}
\begin{split}
\frac{2B'^2}{B^2}+\frac{4B'A'}{AB}-\frac{4B'}{rB}-\frac{4A'}{rA}=0,
\end{split}
\end{equation}

\begin{equation}\label{FE2}
\begin{split}
\frac{-2r^2B''}{B}-\frac{2r^2A''}{A}+\frac{2r^2B'^2}{B^2}-\frac{2rB'}{B}+\frac{4r^2A'^2}{A^2}-\frac{2rA'}{A}=0,
\end{split}
\end{equation}

\begin{equation}\label{FE3}
\begin{split}
\frac{-4BB''}{A^2}+\frac{6B'^2}{A^2}-\frac{8BB'}{rA^2}=0.
\end{split}
\end{equation}
The solution of this system is
\begin{equation}\label{AB}
 A =-c_{1}\left[\frac{c_{2} r-c_{3}}{c_{2} r+c_{3}}\right],\quad  B=\frac{4 r^2}{(c_{2} r - c_{3})^2},
\end{equation}
where $c_{1}$, $c_{2}$ and $c_{3}$ are arbitrary constants. Substituting from (\ref{AB}) into (\ref{cov_met}), the line element for this system will be
\begin{equation}\label{schw1}
ds^2=\left[\frac{c_2 r-c_3}{2 r}\right]^4(dr^2+r^2 d\theta^2+r^2 \sin^2 \theta\, d\phi^2)+\frac{1}{{c_{1}}^{2}}\left[\frac{c_2 r+c_3}{c_2 r-c_3}\right]^2 dt^2.
\end{equation}
Choosing the arbitrary constants as $(c_1, c_2, c_3)=(i,\, 2,\, -m)$, the line element (\ref{schw1}) takes the form

\begin{equation}\label{line_element}
d s^2=\left(1+\frac{m}{2r}\right)^4(dr^2+r^2 d\theta^2+r^2 \sin^2 \theta\, d\phi^2)-\left(\frac{1-m/2r}{1+m/2r}\right)^2 dt^2.
\end{equation}
This gives rise to the proper time $\tau$:
\begin{equation}\label{prop_time}
d \tau^2=-\left(1+\frac{m}{2r}\right)^4(dr^2+r^2 d\theta^2+r^2 \sin^2 \theta\, d\phi^2)+\left(\frac{1-m/2r}{1+m/2r}\right)^2 dt^2,
\end{equation}
which is identical to the Schwarzschild field in isotropic coordinates, $m$ being the geometric mass of the source of the field.

\vspace{5pt}
The linearization of the field equations shows that taking the parameter $b=0$ in Poisson's equation (\ref{poisson1}) gives rise to Laplace's equation,  describing the field in free space. Moreover, the above treatment shows that the field equations with spherical symmetric distribution produces the Schwarzschild exterior solution when considering the matching condition $b=0$. This case corresponds to the code $GI$ according to the type analysis of Table \ref{T4}.

On the other hand, the case $b \neq 0$ may allow the model to describe the field within a material distribution (code $GIV$), which provides an interior solution beneficial in constructing a stellar model. This work is in progress now.


\section{Discussion and Concluding Remarks}\label{S7}
In the present work, we suggest a pure geometric gravity theory. The theory is constructed in the framework of PAP-geometry using the Dolan-McCrea variational method. We conclude the paper by the following comments and remarks.

\begin{itemize}
  \item [1.]
The theory constructed has the following general features:

(\textbf{a})  In $4$-dimensions it has $16$ field equations (\ref{field-eqns}) in $16$-field variables (the tetrad vector field components). All physical quantities covered by the theory are defined in terms of the tetrad vector fields.

(\textbf{b})  A new feature of the theory is that it describes a material-energy distribution in a pure geometric form (\ref{material-energy}).

(\textbf{c})  The theory is generally covariant, i.e., it satisfies the general covariance (relativity) principle.

(\textbf{d})  The PAP-geometry admits a general path equation \cite{W98}
       \begin{equation}\label{par_eqn_motion}
    \frac{d^2 x^\mu}{d \tau^2}+\hat{\Gamma}^{\mu}{_{\alpha \beta}}\,\frac{d x^\alpha}{d \tau}\frac{d x^\beta}{d \tau}=-b\Lambda_{(\alpha \beta)}{^{\mu}}\,\frac{d x^\alpha}{d \tau}\frac{d x^\beta}{d \tau},
\end{equation}
where $\tau$ is a parameter characterizing the path. It is clear that this equation can be used as an equation of motion of any spinning test particle in a gravitational field \cite{W98}. Consequently, the suggested theory violates the weak equivalence principle. This principle is only satisfied in the case $b=~0$, we will discuss this issue in some details latter in this section.

\item [2.]
Three different tools have been used to attribute physical meaning to the geometric objects of the theory. These tools support the theory in a consistent manner.

(\textbf{a})  The first tool (Subsection \ref{S5.1}) is comparing the suggested theory with nonlinear field theories. The symmetric part of the field equations (\ref{symm-field-eqns}) is compared with the field equations of GR. The comparison enables to attribute the gravitational potential to the metric tensor $g_{\mu \nu}$, the field strength to the tensor $G_{\mu \nu}$ and the material-energy distribution to the tensor $\s{T}_{\mu \nu}$. On the other hand, the comparison of the skew part of the field equations (\ref{skew-field-eqns}) with Einstein-Maxwell's theory shows that the theory suggested has no electromagnetic counterpart in the conventional sense. The theory is thus a pure gravitational one.

(\textbf{b})  The second tool (Subsection \ref{S5.2}) is comparing the suggested field theory with linear field theories. For this purpose, the theory has been linearized by assuming a weak static field and a slowly moving test particle. The comparison between the linearized symmetric part of the field equations and Newton's theory supports the above mentioned attributions. Furthermore, a geometric definition of the proper density and proper pressure for an isotropic perfect fluid is proposed and a Poisson-type equation is obtained. The comparison between the linearized skew-part of the field equations and Maxwell's theory supports the above mentioned result that the theory suggested is a pure gravity one.

(\textbf{c})  The third tool, type analysis (Subsection \ref{S5.3}), depends mainly on the expansion Table \ref{T3} and the tensors constituting the field equations. This tool is generally covariant in spite of its dependance on the linearization scheme. It measures the capability of a geometric structure, used in applications, for describing physical fields and gives physical information about this structure even before solving the field equations of the theory.

\item [3.]
Applications of a theory yield more deep physical insight to the geometric objects included in the theory. This is often done in certain special cases characterized by a group of symmetry. The present theory has been applied to a general AP-structure having spherical symmetry.The type of the AP-structure used (\ref{SS-Tetrad}) is found to be $GIV$ and consequently the field equations (\ref{FEb1})-(\ref{FEb3}) are capable of describing a strong gravitational field within a material distribution. The solution in this case needs more deep investigation and is postponed to a future work, in progress now.

\item [4.]
The vanishing of the parameter $b$ of the PAP-geometry reduces the type of (6.1) from $GIV$ to $GI$ and the set of differential equations from (\ref{FEb1})-(\ref{FEb3}) to (\ref{FE1})-(\ref{FE3}). The new type $GI$ indicates that the theory is capable of describing gravitational field outside a spherically symmetric material distribution. Integrating the field equations (\ref{FE1})-(\ref{FE3}) and adjusting the arbitrary constants give rise to the Schwarzschild exterior solution. On the other hand, the equations of motion (\ref{par_eqn_motion}) of the theory reduce, in this case, to the geodesic equations. This permits to conclude that the present theory has the same advantages of GR besides other advantages which may be gained upon relaxing the conditions $b=0$.

\item [5.] The geometry used in the present work contains a parameter $b$, see (\ref{p-canonical}). This is one of the main characteristic of the PAP-geometry. This geometry covers the domain of the Riemannian geometry for $b=0$, while it reduces to the AP-geometry for $b=1$. It is to be noted that the parameter $b$ is neither $0$ nor $1$. The results of three different experiments has been used to fix a value for the parameter $b$ \cite{SM04,WMK2000,G2007}. It is found to be around $10^{-3}$. The theory given in the present work is calculated within the context of the PAP-geometry.

\item [6.] The theory generally violates the WEP. To explore this result in more details, let us start by giving a discussion of the WEP from the point of view of GR, the standard theory for gravity, so far. In this theory there are two types of equations. The field equations
    \begin{equation}\label{GRfield}
        R_{\mu\nu}-\frac{1}{2}g_{\mu\nu}R=-\kappa T_{\mu\nu},
    \end{equation}
    and the geodesic equations of motion
    \begin{equation}\label{GRgeodesic}
        \frac{dU^{\alpha}}{ds}+\overcirc{\Gamma}{^\alpha}{_{\mu\nu}}U^{\mu}U^{\nu}=0.
    \end{equation}
   The WEP is satisfied by the two equations as follows. Due to the minimal coupling used in deriving equation (\ref{GRfield}), the gravitational mass associated with the LHS equals the inertial mass associated with the RHS. The equality of these two masses represents one way of expressing the WEP. The above discussion would not be so clear in the case of the field in empty space, i.e. when the field equations (\ref{GRfield}) becomes
   \begin{equation}\label{GRempty}
    R_{\mu\nu}=0.
   \end{equation}
   So, it is more appropriate to switch the discussion to the equations of motion (\ref{GRgeodesic}). These equations have no parameter related to the intrinsic properties of the moving test particle. The equations of motion (\ref{GRgeodesic}) show that ``free fall is independent of the intrinsic properties of the falling particle". This is an alternative statement of the WEP. It is more appropriate for discussing this principle in any suggested theory.

  Now, let us discuss the WEP in the context of the present work recalling Eddington \cite{Eddington:1923mtr} viewpoint about the WEP
   \begin{quote}
    ``\textit{... it is to be regarded as a suggestion rather than a dogma admitting of no exception.}"
   \end{quote}
   This means that it is not necessary, in general, for this principle to be admitted in the theory. For the present theory, in general, the WEP is not satisfied because of two reasons: The first is the non-minimal coupling between matter and gravity. The second is the RHS of the equations of motion (\ref{par_eqn_motion}) which admits the properties of the moving particle (its spin). But in the case of vanishing $b$ (the spin-torsion coupling), the present theory reduces to GR with all its structure and advantages. In other words, for $b=0$ the field equations (\ref{q42}) and (\ref{electromagnetic}) will be reduced to the GR field equations (\ref{GRempty}). Also, the equations of motion of the theory (\ref{par_eqn_motion}) would reduce to the geodesic equation (\ref{GRgeodesic}). Consequently, in that case, the WEP is satisfied.

\item [7.] Cosmological consequences of the suggested theory need more efforts, in progress now. Using geometric structures, satisfying the cosmological principle, in applications \cite{W86} gives some primary results. The world model obtained is free from particle horizons and flatness problems. More efforts are still needed to explore other properties of this model.
\end{itemize}
%
\end{document}